\RequirePackage{fix-cm}
\documentclass[smallextended]{svjour3}       
\smartqed  
\usepackage{graphicx}
\usepackage{amssymb}
\usepackage[latin1]{inputenc}
\usepackage{hyperref}
\usepackage{amstext}
\usepackage[nearskip,caption=false,font=footnotesize]{subfig}

\usepackage{color}
\definecolor{greencolor}{rgb}{0,0.5,0.2}
\definecolor{redcolor}{rgb}{0,0,0}
\definecolor{bluecolor}{rgb}{0,0.,1.}
\definecolor{greycolor}{rgb}{.5,.5,.5}
\def\Red#1{{\color{redcolor} #1}}

\begin{document}

\title{Topological-collaborative approach for disambiguating authors' names
in collaborative networks
}

\titlerunning{Topological-collaborative approach for name disambiguation}        

\author{Diego R. Amancio
\and Osvaldo N. Oliveira Jr. \and Luciano da F. Costa
}


\institute{Diego R. Amancio \at
Institute of Mathematics and Computer Science \\
University of S\~ao Paulo, P. O. Box 369, Postal Code 13560-970 \\
S\~ao Carlos, S\~ao Paulo, Brazil\\
\\
Osvaldo N. Oliveira Jr. and Luciano da F. Costa \at
S\~ao Carlos Institute of Physics\\
University of S\~ao Paulo, P. O. Box 369, Postal Code 13560-970 \\
S\~ao Carlos, S\~ao Paulo, Brazil\\
}

\date{Received: date / Accepted: date}

\maketitle

\begin{abstract}
Concepts and methods of complex networks have been employed to uncover patterns in a myriad of complex systems. Unfortunately, the relevance and significance of these patterns strongly depends on the reliability of the datasets. In the study of collaboration networks, for instance, \Red{unavoidable noise pervading collaborative  networks} arises when authors share the same name. To address this problem, we derive a hybrid approach based on authors' collaboration patterns and on topological features of collaborative networks. Our results show that the combination of strategies, in most cases, performs better than the traditional approach which disregards topological features. We also show that the main factor accounting for the improvement in the discriminability of homonymous authors is the average shortest path length. Finally, we show that it is possible to predict the weighting associated to each strategy compounding the hybrid system by examining the discrimination obtained from the traditional analysis of collaboration patterns. Because the methodology devised here is generic, our approach is potentially useful to classify many other networked systems governed by complex interactions.
\end{abstract}

\section{Introduction}
\label{intro}

The last few years have witnessed an increasing interest within the scientific  community in the study of the patterns displayed by networked systems such as social networks, the Internet, the WWW and biological structures~\cite{applications}. The study of social networks, for instance, has been useful to unveil many properties of collective interactions such as social influence~\cite{social1,social2,social3}, disease spreading~\cite{epidemic1,epidemic2,epidemic3} and many other sociologic aspects. Of particular interest to the aims of this study are the collaborative networks, which are among the largest social networks ever studied~\cite{scollaboration}. The main findings achieved with the modeling of social interactions as collaborative networks include the observation of non-trivial topological information such as the small-world phenomena~\cite{swp}. This means that, similarly to random graph models~\cite{random}, the typical distance between nodes scales logarithmically with the number of agents in the system. Another interesting pattern in the counterpart network, referred to as citation network~\cite{citnet}, is the cumulative behavior~\cite{simon57,price76}, whereby papers referenced on many occasions tend to be cited again more intensely than other less referenced papers, thus reflecting the preferential attachment (or richer-get-richer paradigm)~\cite{barabasi99}.

Collaborative networks have also been employed in information sciences to disambiguate authors' names in scientific manuscripts~\cite{malin}. Among many reasons, the accurate identification of authorship in manuscripts plays a fundamental role in the scientific community. For example, the \Red{imprecise} quantification of researchers' merit based on their publication profile might lead to unfair decisions regarding the distribution of project financing, awards and so forth, thus undermining the efficiency of the system as a whole. Despite the many attempts to handle the task satisfactory~\cite{surveyAqui}, a robust, precise system has not been conceived yet. For this reason, the disambiguation task remains an unsolved task for information sciences. Traditional network methods for addressing the problem relies on the assumption that homonymous authors can be distinguished by examining the distinct patterns of collaboration~\cite{malin}, because they often collaborate with different colleagues.
%
Unfortunately, some relevant network knowledge has been widely disregarded. For example, traditional approaches do not take into account the topological information of collaborative networks:
patterns of connection with immediate neighbors have not been considered. In the same way, the information concerning the global connectivity of authors in their scientific community have also been neglected. In this paper, we analyze the suitability of topological measurements as a complementary feature for the authors' name disambiguation task through the \Red{conception of a hybrid system compounding both traditional and topological strategies.} More specifically, we show that the ability of discrimination is improved in three data sets with the proposed convex combination of classifiers. \Red{We also found that the most relevant topological feature for discriminating authors' names is the average shortest path length}, which suggests that the long-range connectivity patterns plays a fundamental role in characterizing authors unequivocally. Finally, we show that the weighing associated with the topological strategy can be obtained automatically \Red{because it strongly correlates} with the discriminability obtained with the traditional approach.

This paper is organized as follows. In Section \ref{metodologia}, we detail the methodology employed to combine pattern recognition techniques. In the same section, we introduce the model based on collaborative networks and the topological features employed to characterize the organization of these networks. In Section \ref{resultados}, we display the results and discussions obtained with the proposed technique. Finally, in Section \ref{conclusao}, \Red{we conclude the manuscript and suggest future investigations}.

\section{Traditional approaches for authors' name disambiguation}

{
The discrimination of authors' names can be defined as the ability to determine which individual authoring a document is behind an alias. To define the task, let $\mathcal{D}$ be a set of documents in a dataset. Consider $\mathcal{N}$ as the set of aliases (authors' names) in $\mathcal{N}$ and
$\mathcal{A}$ the set of authors. We aim at creating a mapping $\mathcal{M} : \mathcal{N} \mapsto \mathcal{A}$, where $\mathcal{M}$ is not necessary an injective function.
The mapping $\mathcal{M}$ is supposed to assign more than one individual to each alias in a given document $d \in \mathcal{D}$, although typically only one is chosen.
}

{
Names inaccuracies represent a major obstacle in processing and interpreting large \Red{amounts} of scientific papers for the purpose of grants and promotions. Because many applications require data devoid of noise
~\cite{clean}, several methods have been developed to identify and remove ambiguities at the paper level. The simplest techniques are those based solely on authors' names. These methods can either use the first or all initials to represent a persona~\cite{scollaboration}. More refined methods use additional information to perform the discrimination of ambiguous names. The most common complementary features currently employed are the list of references~\cite{initials}, textual content~\cite{huang}, crowd-sourced intelligence~\cite{crowding}, citation counts~\cite{tang}, co-authors' names~\cite{ongraph} and publication venue title~\cite{venue}. Interestingly, even techniques borrowed from other research fields such as the use of stylometry features have been proposed to deal with ambiguities~\cite{surveyAqui}.
}

{
Of paramount importance to the purposes of the current study are the graph-based techniques, which basically \Red{represent the collaborations between authors as edges of complex networks}~\cite{malin,amancio1,timevarying}. The most traditional graph-based strategies
assume that authors tend to maintain its collaboration group. As a consequence, distinct personas can be related to each other through the analysis of recurrence of neighbors~\cite{malin}. In others words, if two personas share many neighbors (collaborators) in the collaborative network, then they probably are the same author. The above \Red{strategy has been} generalized to include not only immediate neighbors in the analysis, but also neighbors of neighbors and further hierarchies~\cite{amancio1}.
}

{
The reference~\cite{quasicliques} deals with a task related to the discrimination of homonymous authors. More specifically, they investigate how to find variations of authors' names in publications.
Taking as premise the fact that authors tend to maintain a context
(e.g. co-authors' names or words employed in paper titles), they perform an unsupervised classification. At the core of their method, graph similarity measurements are combined with textual techniques to cluster distinct variations. An interesting contribution was proposed in~\cite{ongraph}, which also employs a pre-processed collaborative network model devoid of invalid, redundant paths. Upon using a similarity measurement inspired by the traditional electric circuit theory, they cluster ambiguous instances according to the Affinity Propagation criteria~\cite{afinidade}. A different collaborative network is proposed in~\cite{redediff}. In such a network,  nodes might represent either papers or authors. As for the edges, there is a twofold definition: they either connect authors sharing the same alias or papers and their respective authors. Through the definition of a set of match functions, the proposed technique combines social network analysis with traditional similarity functions, yielding a significant improvement in the performance when compared with traditional techniques disregarding the network nature of scientific interactions.
}

\Red{
As noted above, the graph representation is a very common representation for disambiguating authors' names. Nevertheless, the topological information has been widely neglected. The ability of the topological properties of collaborative networks to discriminate authors was first observed in~\cite{amancio1}. In such study, it was shown that it is possible to disambiguate homonymous authors using topological measurements of complex networks. The best results, however, were achieved with further hierarchies of traditional collaborative attributes. In the current study, we extend the work conducted in~\cite{amancio1} by devising a hybrid strategy that uses topological features to improve the traditional approach. More specifically, we show that the topological factor can be employed as decision factor whenever the traditional network features (e.g. the recurrence of co-authors) is not useful to assign the correct author to an alias under dispute.}

\section{Topological-collaborative approach} \label{metodologia}

{The use of topological measurements to \Red{characterize topological patterns in collaborative networks} was motivated by the use of similar strategies in the characterization of complex systems~\cite{surveymeas}. \Red{It has been shown that collaborative networks exhibit a modular organization} with well-defined collaborative groups or communities~\cite{timevarying}. Because \Red{nodes belonging to the same communities tend to share certain topological features~\cite{eigen}}, it is likely that nodes belonging to distinct communities will discriminate distinct authors.}

The strategies employed in this paper to disambiguate authors' names are based in the construction of collaborative networks. In Section \ref{netfom} we show how collaborative networks representing the relationship between authors are built. Then the nodes representing ambiguous names are characterized with traditional (or collaborative) features and with attributes extracted from topological measurements of complex networks (CN)~\cite{surveymeas} (see Section \ref{caracterizacao}). The methodology employed to combine both approaches in a hybrid strategy is then presented in Section \ref{sec:fuzzy}.

\subsection{Network formation} \label{netfom}

A network $\mathcal{G}$ is defined as $\mathcal{G} = \{\mathcal{V},\mathcal{E}\}$, where $\mathcal{V}$ represents the set of nodes and $\mathcal{E}$ represents the set of edges. In a collaborative network, nodes represent authors. An edge exists between two authors if they published at least one paper together. Each homonymous author is represented as a different node because \Red{the disambiguation is performed at the paper level~\cite{malin,amancio1}, i.e. one alias at a time is disambiguated}.

\begin{figure}
    \begin{center}
        \includegraphics[width=0.6\textwidth]{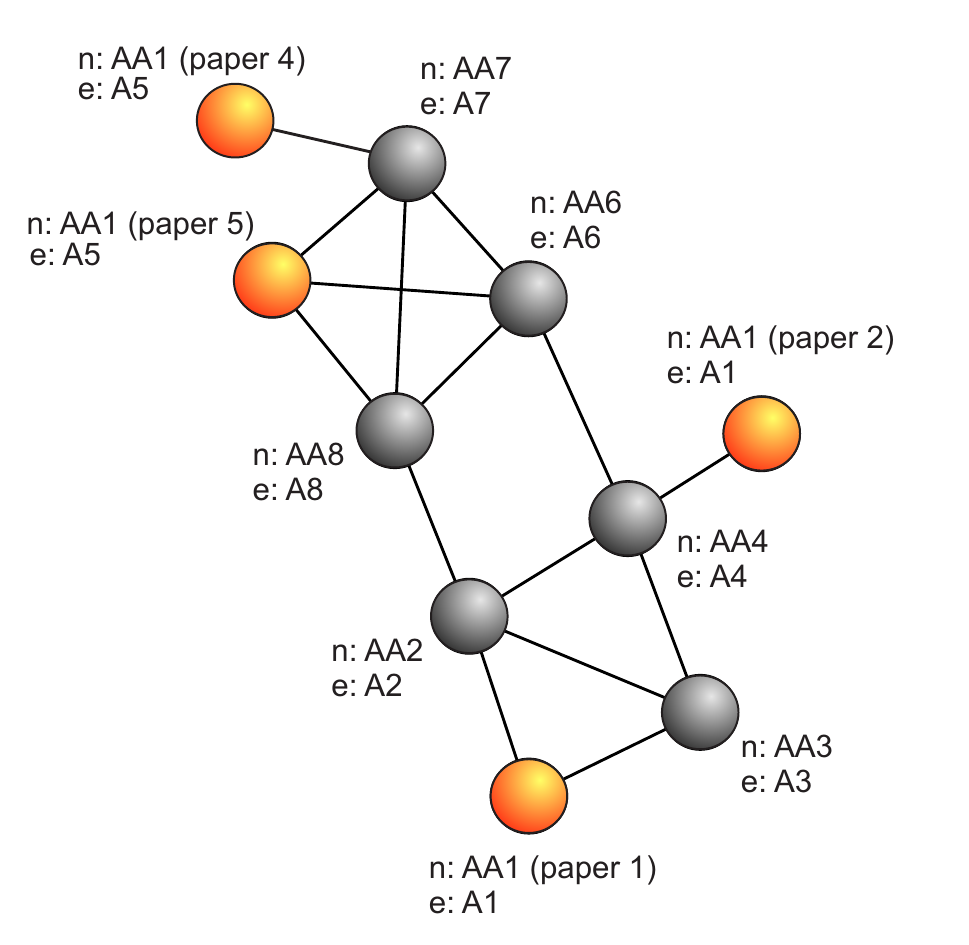}
    \end{center}
    \caption{\label{fig:network_example}\Red{Example of collaborative network obtained for the following toy database: Paper 1 ({AA1}, {AA2} and {AA3}), Paper 2 ({AA1} and {AA4}), Paper 3 ({AA3}, {AA4} and {AA2}), Paper 4 ({AA1} and {AA7}), Paper 5 ({AA1}, {AA6}, {AA7} and {AA8}), Paper 6 ({AA6} and {AA4}) and Paper 7 ({AA8} and {AA2}). In the figure, \emph{e} represents the entity (author) and \emph{n} represents the alias (name) used to represent the entity in the paper. Since authors {A1} and {A5} share the same alias {AA1}, in this example, each of their occurrences is presented as a different node in the network.}
    }
\end{figure}

The strength of the link connecting nodes $i$ and $j$ is
\begin{equation}
w_{ij} = \sum_{\rho \in \Pi} \frac{ \delta_{ij\rho} }{ \|\rho\| }, \; \; 
\label{eq:formulaDiego}
\end{equation}
where $\Pi$ is the set comprising all papers in the database, $\|\rho\|$ represents the number of authors in paper $\rho$ and
\begin{equation} \label{cond}
\delta_{ij\rho} = \left\{
\begin{array}{rl}
1 & \textrm{if  $i$ and $j$ collaborate in paper $\rho$,} \\
0 & \textrm{otherwise. }\\
\end{array} \right.
\end{equation}
In eq. (\ref{eq:formulaDiego}), $\|\rho\|$ is employed as a normalization factor to take into account the fact that the strength of links connecting a few authors tend to be more intense than those involving many authors~\cite{malin}. \Red{To illustrate the process of building the network, consider a toy database comprising $7$ papers (paper 1-7), $8$ authors (A1-8) and 7 distinct aliases (AA1, AA2, AA3, AA4, AA6, AA7 and AA8), as shown in Table \ref{tabela1}. The network obtained from the example is illustrated in Figure \ref{fig:network_example}. We hypothesize that authors {A1} and {A5} share the same alias AA1}. For this reason, each {apparition (persona) is}
considered as a distinct node in the network (see highlighted nodes). According to eq. (\ref{eq:formulaDiego}), the strength $w_{12}$ of the edge linking {AA1} and {AA2} is 1/3 (from paper 1). In a similar fashion, $w_{23} = 2/3$, i.e., 1/3 (from paper 1) plus 1/3 (from paper 3).

\begin{table}
\centering
\caption{\label{tabela1}Authorship in a toy dataset comprising 7 papers. Names (n) represent the list of names as it appears in the paper. Entities (e) represent the author assigned to the respective names. In the example, AA1 is an ambiguous name because it names both authors A1 and A5.}
\begin{tabular}{|l|l|l|}
  \hline
  {\bf Papers} & {\bf Names (n)} & {\bf Entities (e)} \\
  \hline
  Paper 1 & AA1, AA2 and AA3 & A1, A2 and A3 \\
  Paper 2 & AA1 and AA4 &  A1 and A4\\
  Paper 3 & AA3, AA4 and AA2 & A3, A4 and A2 \\
  Paper 4 & AA1 and AA7 & A5 and A7 \\
  Paper 5 & AA1, AA6, AA7 and AA8  & A5, A6, A7 and A8 \\
  Paper 6 & AA6 and AA4 & A6 and A4 \\
  Paper 7 & AA8 and AA2 & A8 and A2 \\
  \hline
\end{tabular}
\end{table}

\subsection{Network characterization} \label{caracterizacao}

The characterization of each node was performed in two different ways. In the first strategy, referred to as \emph{collaborative} method, nodes representing ambiguous names are characterized by the weight $w_{ij}$ of their respective edges.
In other words, the node $v_i$ is characterized by the vector  $\vec{\bf w}(i) = (w_{i1}, w_{i2} \ldots w_{in})^T$, where $n = \| \mathcal{V} \|$ is the total number of authors. Thus, if two ambiguous nodes tend to connect to the same neighbors
one expects that $\| \Delta \vec{\bf w} \| \rightarrow 0$. In the second strategy, referred to as \emph{topological} approach, nodes are represented mathematically with the vector $\vec{\bf \mu}(i) = (\mu_{i1}, \mu_{i2} \ldots \mu_{iF})$ storing $F$ topological measurements of complex networks~\cite{surveymeas}. This strategy is based on the conjecture that the same author tends to display a regular organization in the \Red{collaborative network} so that the connectivity patterns of distinct authors might be differentiated by the number of links, by the average connectivity of neighbors and so on. In the current study, the following CN measurements were employed:

\begin{itemize}

  \item {\bf Degree}: the degree $k_i^{(1)} = \sum_j a_{ij} = \sum_k a_{ki}$, where
        \begin{equation} \label{cond}
a_{ij} = \left\{
\begin{array}{rl}
1 & \textrm{whenever  $w_{ij} > 0$,} \\
0 & \textrm{otherwise }\\
\end{array} \right.
\end{equation}
quantifies the number of neighbors connected to the node $v_i$. {In other words, it quantifies the \Red{total number} of different \Red{co-authors.}}.

\item {\bf Strength}: represents the sum of the edges weights of the neighbors and is computed as $s_i^{(1)} = \sum_j w_{ij} = \sum_k w_{ki}$.

  \item {\bf Neighborhood degree}: the neighborhood degree distribution of $v_i$ is summarized with the average $k_{i,n}$
       \begin{equation}
            k_{i,n} = \sum_{j} a_{ij} k^{(1)}_i  / \sum_{j} a_{ij}
       \end{equation}
       and with the standard deviation $\Delta k_{i,n}$
       \begin{equation}
            \Delta k_{i,n} = \Bigg{[} \sum_{j} a_{ij} \Big{(} k_j^{(1)} - \sum_{k} a_{ik} k^{(1)}_i  / \sum_{k} a_{ik} \Big{)} ^ 2 / \sum_{j} a_{ij} \Bigg{]}^{1/2}.
       \end{equation}
       {Both measurements quantify the collaboration patterns of the authors that collaborated at least once with the author being analyzed.}

  \item {\bf Neighborhood strength}: the weighted connectivity of the neighbors of $v_i$ is represented with the average $s_{i,n}$
      \begin{equation}
        s_{i,n} = \sum_{j} a_{ij} s^{(1)}_i  / \sum_{j} a_{ij}
      \end{equation}
      and standard deviation $\Delta s_{i,n}$
      \begin{equation}
          \Delta s_{i,n} = \Bigg{[} \sum_{j} a_{ij} \Big{(} s_j^{(1)} - \sum_{k} a_{ik} s^{(1)}_i  / \sum_{k} a_{ik} \Big{)} ^ 2 / \sum_{j} a_{ij} \Bigg{]}^{1/2}
      \end{equation}
      of the strength of the neighbors.

   \item {\bf Clustering coefficient}: the quasi-local topological structure of nodes is measured in terms of the clustering coefficient $C^{(1)}$, which quantifies the extent to which neighbors tend to cluster together. {More specifically, the clustering coefficient quantifies
       the fraction of collaborations that are also established among collaborators}.
       The local version of this measurement quantifies how close are the neighbors to become a clique, i.e., a fully connected subgraph. Using the adjacency matrix, $C^{(1)}$ is computed as
       \begin{equation} \label{aglomeracao}
            C^{(1)}_i = 3 \sum_{k > j > i} a_{ij} a_{ik} a_{jk}  \Bigg{[} \sum_{k > j > i} a_{ij} a_{ik} + a_{ji} a_{jk} + a_{ki} a_{kj} \Bigg{]} ^ {-1}.
       \end{equation}

    \item {\bf Average shortest path length}: this measurement is calculated from the quantity $d_{ij}$ measuring the minimum cost required to reach a given node $v_j$: 
        \begin{equation}
            l_i = \frac{1}{ n - 1} \sum_{i=1}^{n} d_{ij}.
        \end{equation}
        {Unlike the above measurements, the average shortest path length analyzes the global patterns of connectivity of authors}. \Red{This measurement is only calculated for author instances belonging to the same component}.

    \item {\bf Betweenness}: the centrality of nodes was also examined in terms of the total the number of shortest paths that pass through them. If $\sigma(v_i,v_u,v_j)$ represents the number of shortest paths from node $v_i$ to node $v_j$ passing through $v_u$, and $\sigma(v_i,v_j)$ is the total number of shortest paths from $v_i$ to $v_j$, then the betweenness $B$ is computed as:
        \begin{equation} \label{obtw}
            B_u = \frac{\sigma(v_i,v_u,v_j)}{\sigma(v_i,v_j)}.
        \end{equation}
        The betweenness is useful to distinguish, p.e. leading authors from the others because it is a centrality measurement. As such, it assigns a relevance score for each author according to eq.~(\ref{obtw}). Unlike other centrality measurements (degree and strength), it does not disregard the global connectivity of the network to quantify the centrality of authors. \Red{Similarly to the average shortest path length, this measurement is only calculated for author instances belonging to the same component.}

    \item {\bf Hierarchical measurements}: the hierarchical representation of traditional measurements takes into account further neighborhoods to quantify the local structure of nodes~\cite{concentric,hierarchical}. To define this level of representation, consider the following definitions. Consider that $\mathcal{A}$ and $\mathcal{W}$ are the matrices storing the elements $a_{ij}$ and $w_{ij}$, respectively.
        Let $\mathcal{R}_1(i)$ be the set of immediate neighbors of $v_i$. This set accounts for the transmission of information from $v_i$ to every node belonging to the set $\{v_j \in \mathcal{V}~|~v_j \notin \mathcal{R}_1(i)\}$. Hence, $\mathcal{R}_1(i)$ is responsible for the creation of ``virtual edges'' connecting $v_i$ and the nodes two hops away from $v_i$. Mathematically, $\mathcal{R}_1(i)$ is deduced from the vector $\mathbf{\vec{\nu}_1(i)}$:
        \begin{equation}
			\mathbf{\vec{\nu}_1(i)} = \mathcal{A} \cdot \vec{\tau}(i), \nonumber
        \end{equation}
        where
\begin{equation} \label{cond}
\tau_j(i) = \left\{
\begin{array}{rl}
1 & \textrm{whenever  $i = j$,} \\
0 & \textrm{otherwise. }\\
\end{array} \right.
\end{equation}

The quantity $\vec{\rho_1}(i)$ is defined with the Heaviside function $\delta$ as $\vec{\rho_1}(i) = \delta ( \vec{\nu}_1(i) )$. Thus, if $v_j \in \mathcal{R}_1(i)$, then $\vec{\rho}_1(i) = 1$. Otherwise, $\vec{\rho}_1(i) = 0$. The definition of $\vec{\nu}_1$($i$) can be extended to consider further hierarchies. This is achieved with the vector $\vec{\nu}_h$($i$), whose non-zero elements represent the set $\{v_j \in \mathcal{V}~|~d_{ij}=h\}$, i.e., $\mathcal{R}_h$:
\begin{equation}
			\vec{\nu}_h(i)  = \mathcal{A}^h \cdot \vec{\tau}(i) \label{vvv}. 
\end{equation}
Analogously to the definition of $\vec{\rho_1}$, $\vec{\rho_h}$ \Red{is defined} as the set of nodes whose maximum distance from node $v_i$ does not exceed $h$ steps:
\begin{equation} \label{novaequacao}
    \vec{\rho_h}(i) = \delta \Bigg{(}  \sum_{k=1}^{h} \vec{\rho_k}(i) + \vec{\nu}(i) \Bigg{)}. \label{xxx}
\end{equation}
Note that, with the definition of $\vec{\rho_h}$ in eq. (\ref{novaequacao}), $\vec{\nu}_h$ might be computed alternatively as
$\vec{\nu}_h(i) = \vec{\rho_h}(i) - \vec{\rho}_{h-1}(i)$.
Let $\gamma_h$ the subnetwork comprising nodes inside the radius $\mathcal{R}_h(i)$ plus the edges between the nodes in $\mathcal{R}_h(i)$. The hierarchical level $h$ of the network is defined as $\gamma_h$ plus the edges going to $\gamma_{h+1}$. Thus, the hierarchical versions of the traditional measurements are computed by collapsing the nodes inside the radius $\mathcal{R}_h(i)$ as a single node, disregarding both internal nodes and edges. An example of hierarchy is illustrated in Figure \ref{figmotivos}. In the current study, we employed the second and third hierarchies for the degree ($k^{(2)}$ and $k^{(3)}$), strength ($s^{(2)}$ and $s^{(3)}$) and clustering coefficient ($C^{(2)}$ and $C^{(3)}$).
			
\begin{figure}
    \begin{center}
        \includegraphics[width=0.8\textwidth]{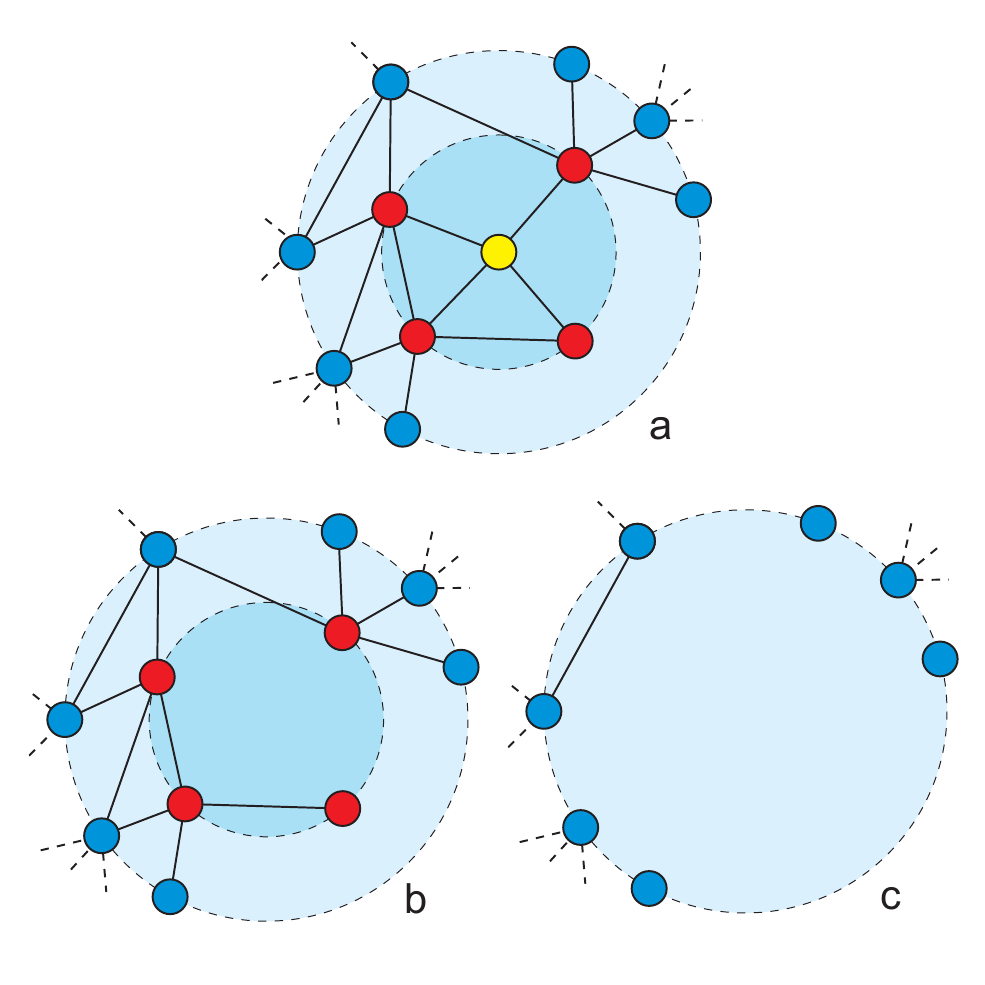}
    \end{center}
    \caption{Examples of hierarchies calculated for the yellow central node depicted in network (a). Red nodes belong to the set $\mathcal{R}_1(\textrm{yellow})$, while blue nodes belong to set $\mathcal{R}_2(\textrm{yellow})$. In order to compute the hierarchical measurements, nodes belonging to a given hierarchy (i.e., nodes inside the circles delimited by dotted lines) are collapsed into a single node, while external (dotted) edges are kept. In panel (b), the neighbors of the yellow node were collapsed into a single node, while in panel (c) the neighbors of the neighbors were collapsed into a single node. Hence, the hierarchical degrees at the second and third levels are respectively $k_{yellow}^{(2)} = 7$ (there are seven blue nodes) and $k_{yellow}^{(3)} = 9$ \Red{(there are 9 edges leaving blue nodes)}. The other hierarchical measurements can be computed in an analogous manner.
    }
    \label{figmotivos}
\end{figure}

\item {{\bf Locality index}:  this quantity measures the proportion of links located at the first hierarchy as:
    \begin{equation}
        \mathcal{L}_i = \frac{k_i^{(1)}}{k_i^{(1)}+k_i^{(2)}}.
    \end{equation}
    }

\end{itemize}

\subsection{Fuzzy algorithm description} \label{sec:fuzzy}

The pattern recognition technique we use is an extension of the well-established ``crisp'' $\kappa$NN algorithm~\cite{duda}, an algorithm that has been shown to be both computationally efficient and highly effective in medium scale datasets~\cite{fuzzy1,fuzzy2}.
The ``crisp'' $\kappa$NN algorithm  is defined as follows. Consider a training dataset $\mathcal{X} = \{x_1,x_2,x_3, \ldots , x_N \}$. As we shall show, this set will be used to detect useful patterns for the disambiguating task. The class\footnote{{The class is the correct name associated to each persona.}} $i$ of each $x_k$, $1 \leq k \leq N$ is known beforehand. Given a new unclassified object $x$, one desires to classify it in the most likely class. To achieve this, the algorithm initially selects the $\kappa$ nearest objects of the training dataset. Then, the most frequent class in this $\kappa$-set is used to decide the class of $x$. If a tie occurs for two or more classes, the algorithm selects the class of the nearest object belonging to one of the tied classes. The obvious limitation of the ``crisp'' $\kappa$NN algorithm is that the training objects pertaining to the same class are equally representative for that class. This deficiency is evident when one compares objects situated near the centroid against those located at the border of the class.
Another problem concerning the algorithm is that $x$ is supposed to belong to a specific class alone.
There is no concept of \Red{membership strength}~\cite{duda}. To overcome such problems, we derived a different approach based on the fuzzy definition proposed in~\cite{fuzzy2}.

To define the fuzzy algorithm, consider the following definitions. Let $u_{ij} \in [0,1]$ be the  membership strength of $x_j$ in class $i$. By definition, $u_{ij}$ is restricted to the following constraints:
\begin{equation} \label{representacao}
    \sum_{i=1}^c u_{ik} = 1,
\end{equation}
\begin{equation} \label{classes}
    0 < \sum_{k} u_{ik} < N,
\end{equation}
where $c$ is the number of classes. Eq. (\ref{representacao}) allows us to interpret $u_{ik}$ as the likelihood of $x_k$ to belong to class $i$. Therefore, if $c=2$ and $u_{1k} = 0.95$ then it is very likely that $x_k$ belongs to $i=1$. Differently, if $u_{1k} = 0.55$ then $x_k$ \Red{is probably  located} at the border of that class. Eq. (\ref{classes}) simply imposes that each class cannot contain more than $N$ objects. In the fuzzy algorithm, the non-normalized membership strength $\tilde{u}_i(x)$ of an unclassified object $x$ in the $i$-th class
is computed \Red{as a function} of the distance of $x$ to the $\kappa$ nearest neighbors:
\begin{equation}
\tilde{u}_i(x) = \sum_{j=1}^{\kappa} u_{ij} \| x - x_j \| ^ {2 / (1-m)}.
\end{equation}
Note that $\tilde{u}_i(x)$ depends on both the distance to the nearest $\kappa$ objects and the \Red{membership strength} $u_{ij}$ of the training set. As such, $\tilde{u}_i(x)$ will take large values whenever the distance between $x$ and $x_j$ is small and $x_j$ is a strongly representative object in class $i$ (i.e., $u_{ij}$ is large). The parameter $m$ accounts for the weight associated to the distance
$\| x - x_j \|$. If $m$ \Red{takes high values}, $\| x - x_j \|$ weakly interferes on the the computation of $\tilde{u}_i(x)$. On the other hand, whenever $m \rightarrow 1$ the influence of the nearest objects becomes so strong that the contribution of the farthest objects in the $\kappa$-set practically disappears. To assure that $u_i(x)$ represents a probability, it is necessary to normalize $\tilde{u}_i(x)$ so that $\sum_i \alpha \tilde{u}_i(x) = 1$. As such, $\alpha$ can be computed from
\begin{equation}
\begin{array}{rcl}
     \alpha^{-1}   & = &    \sum_i  \tilde{u}_i(x)  \\
              & = &  \sum_i \sum_j u_{ij} \| x - x_j \| ^ {2 / (1-m)}  \\
              & = &  \sum_j \Big{[} \sum_i u_{ij} \| x - x_j \| ^ {2 / (1-m)} \Big{]}  \\
              & = &  \sum_j \| x - x_j \| ^ {2 / (1-m)} \sum_i u_{ij}  \\
              & = & \sum_j \| x - x_j \| ^ {2 / (1-m)}.
\end{array}
\end{equation}
Therefore, $u_i(x)$ is given by:
\begin{equation} \label{membresia}
    u_i(x) = \frac{\sum_{j=1}^{K} u_{ij} \| x - x_j \| ^ {2 / (1-m)}}{\sum_{j=1}^{K} \| x - x_j \| ^ {2 / (1-m)}}.
\end{equation}

\begin{figure}
    \begin{center}
        \includegraphics[width=0.35\textwidth]{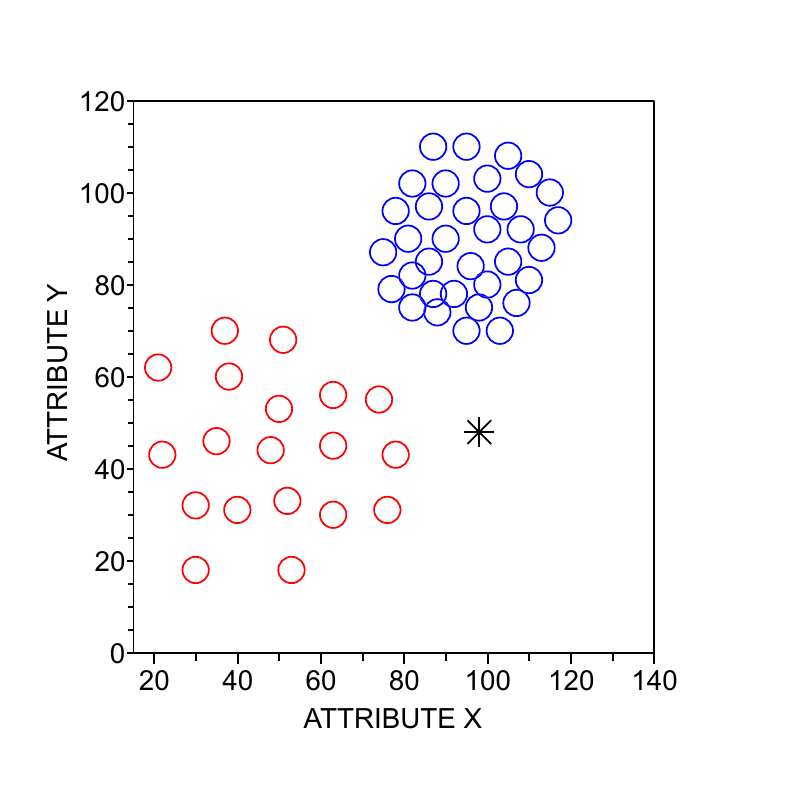}
    \end{center}
    \caption{\label{fig:maha}$d^{(euc)}$($p$,red) / $d^{(euc)}$($p$,blue) = 1.15 and $d^{(mah)}$($p$,red) / $d^{(mah)}$($p$,blue) = 0.62. Even though the Euclidian distance from $p$ to the two groups are similar, $p$ is more similar to the red group according to the Mahalanobis distance because the latter is \Red{sparser}.}
\end{figure}

In order to define the strength $u_{ij}$ of each object $x_j$ in the training set with regard to class $i$, consider the following definitions. $\mu_i = \sum_{j, x_j \in i} x_j / n_i$ represents the centroid of class $i$ comprising $n_i$ objects and cov$(a,b)$ is the covariance between attributes $a$ and $b$:
\begin{equation}
  \textrm{cov}(a,b) = \langle (a - \langle a \rangle )(b - \langle b \rangle) \rangle,
\end{equation}
where $\langle \ldots \rangle$ is the average over all observations of the variables. The covariance matrix between all attributes is defined as $S = [\textrm{cov}(a,b)]$. The similarity between object $x_j$ and a class $i$ is given by
\begin{equation}
     \sigma(i,x_j) = \Big{[} (x_j - \mu_i)^T S_i^{-1} (x_j - \mu_i) \Big{]} ^ {-1/2},
\end{equation}
which represents the inverse of the Mahalanobis distance~\cite{duda}. In particular, such similarity measurement was chosen because it takes into account the dispersion of the classes to compute the distance between $x_j$ and the cluster $i$. In this way, if $x_j$ is far away from the centroid $c_i$ but is included inside the dispersion region of $i$, \Red{then the distance} between $x_j$  and $i$ will be small. Conversely, if $x_j$ is close to $c_i$ but the dispersion of $i$ is too small so that $x_j$ falls outside the boundaries of $i$, then then distance between $x_j$  and $i$ will be probably greater than the previous scenario. {This effect is illustrated in Figure \ref{fig:maha}}.

Finally, the \Red{membership strength} of $x_j$ representing a training object in class $i$ is given by
\begin{equation}
    u_{ij} = \frac{\sigma(i,x_j)}{\sum_i \sigma(i,x_j)}.
\end{equation}
Note that $\sigma(i,x_j)$ is normalized by the sum of the similarity between $x_j$ and all classes in order to satisfy eq. (\ref{representacao}).

\subsection{Hybrid classification and evaluation}

The fuzzy algorithm described in Section \ref{sec:fuzzy} is employed to construct the hybrid classifier as follows. If $u^{(T)}_{ij}$ represents the  membership strength of the object $x_j$ in class $i$ when the collaborative attributes are employed and $u^{(C)}_{ij}$ represents the strength when the CN topological measurements are used, then the membership strength of $x_j$ in class $i$ in the hybrid approach is
\begin{equation} \label{lambida}
    u^{(H)}_{ij} = \lambda u^{(T)}_{ij} + (1-\lambda) u^{(C)}_{ij}, \ \ \ \textrm{$\{\lambda \in \mathbb{R}~|~0 \leq \lambda \leq 1\}$}.
\end{equation}
Note that $\lambda$ in eq. (\ref{lambida}) plays the role of weighting distinct strategies. The larger is the value of $\lambda$, the greater is the importance ascribed to the topological strategy and vice-versa. The class $i$ associated to object $x_j$ is the one satisfying the relation
\begin{equation}
    \hat{u}_{ij}^{(H)} = \max_i u_{ij}^{(H)}.
\end{equation}

Many evaluation methods have been employed to compare the performance of classifiers~\cite{duda}. For simplicity's sake, we measure the distinguishability of homonymous authors as the accuracy rate (see below) provided by the 10-fold cross-validation strategy~\cite{duda}. This strategy is robust because it can be used even when there is not a selected set of instances to assess the learning quality. In this method, the training set is randomly separated into 10 parts so that 9 parts are used for training and the remaining one is employed for evaluation. The process is iterated 10 times until all parts have been used for evaluation. Finally, the accuracy rate is computed as the average accuracy rate obtained along this iterative process.

\Red{Following the data-mining literature~\cite{comparacoes,Witten:2005}, we chose the measurement referred to as accuracy rate to evaluate the disambiguation performance. In each iteration of the cross validation, the confusion matrix $\mathcal{Z}_{n x m}$ is employed in the evaluation. Each element $z_{ij}$ of $\mathcal{Z}$ represents the number of times that the classifier classified author $i$ as being author $j$. Thus, the accuracy rate $\Gamma$ for each iteration is given by
\begin{equation}
    \Gamma = \sum_i z_{ii} \Big{/} \sum_i \sum_j z_{ij},
\end{equation}
because the diagonal elements quantifies the sum of instances
whose author assigned by the algorithm is the correct choice.
Note that this measurement captures errors arising from author splitting, which occurs when the method decides that a given alias belongs to $m$ authors when in reality it belong to $n$, $n<m$. The total amount of author splitting errors is given by
\begin{equation}
    \epsilon_{s} = \sum_{i=1}^{n} \sum_{j=n+1}^m z_{ij}.
\end{equation}
}

\section{Results and discussion} \label{resultados}

\begin{figure}
    \begin{center}
        \includegraphics[width=0.8\textwidth]{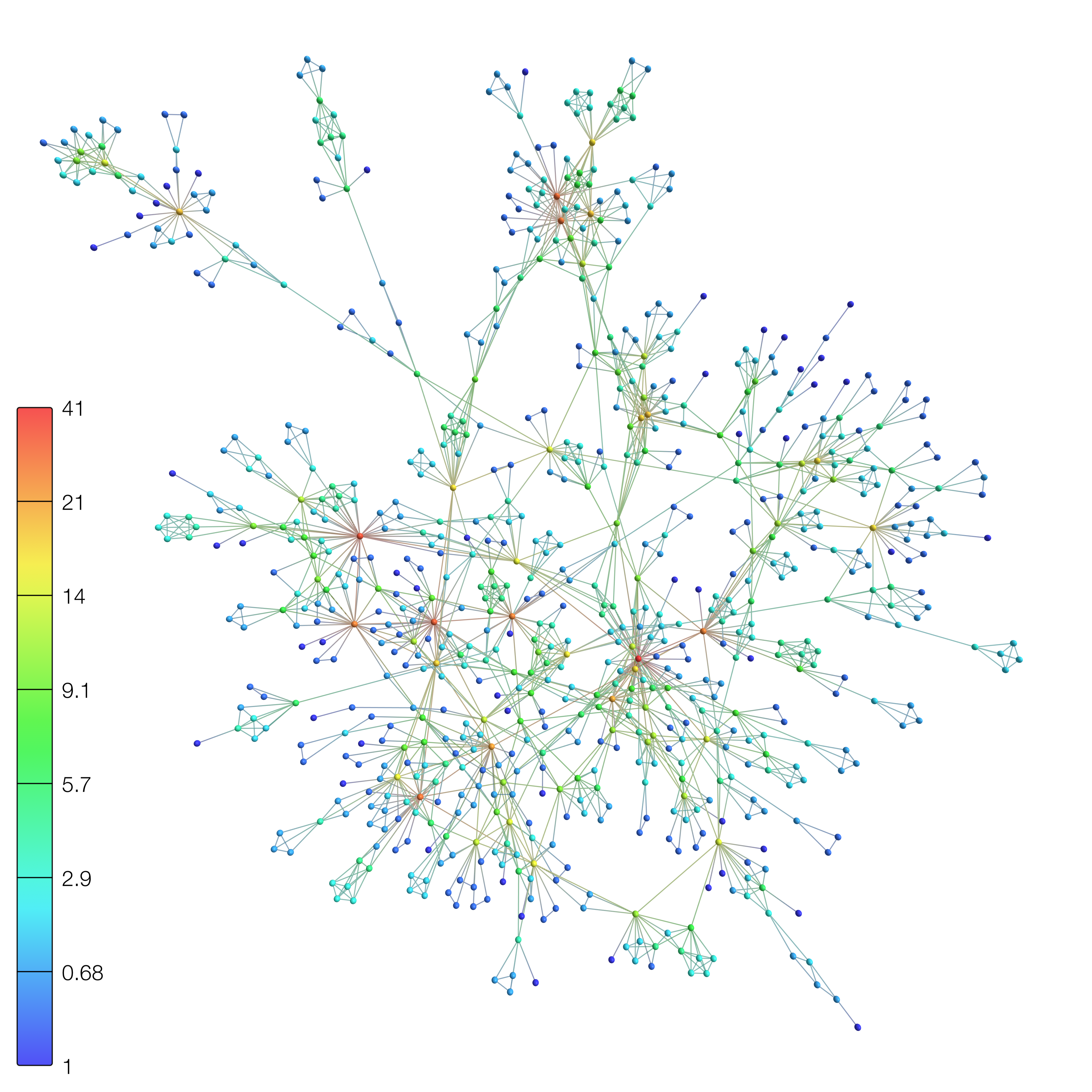}
    \end{center}
    \caption{\label{fig:network}Giant component of the collaborative network obtained from the CN database. Node colors are proportional to the total number of distinct collaborators.}
\end{figure}

In this section, we describe the results obtained with three data sets derived from the arXiv repository, which have been used in previous studies~\cite{amancioInfor,timevarying}.
{The three collaborative networks comprise papers from the following research areas: ``complex networks'', ``topological insulators'' and ``liquid crystals'', which are henceforth abbreviated as \textbf{CN}, \textbf{TI} and \textbf{LC}, respectively. \Red{The papers extracted for CN, TI and LC were published between the years 1999-2013, 2004-2013 and 1993-2013, respectively.}
The papers in each area were recovered from a search for representative keywords appearing in the title or abstract.  In the arXiv database, most of the ambiguities occur due to names inconsistencies. To form the gold standard, we manually disambiguated all inconsistent apparitions.}
The CN collaborative network is illustrated in Figure \ref{fig:network}. In all experiments, we have employed $m=2$ in eq. (\ref{membresia}).

\subsection{Comparing the hybrid with the traditional approach} \label{comparacao}

\begin{figure}
    \begin{center}
        \includegraphics[width=1\textwidth]{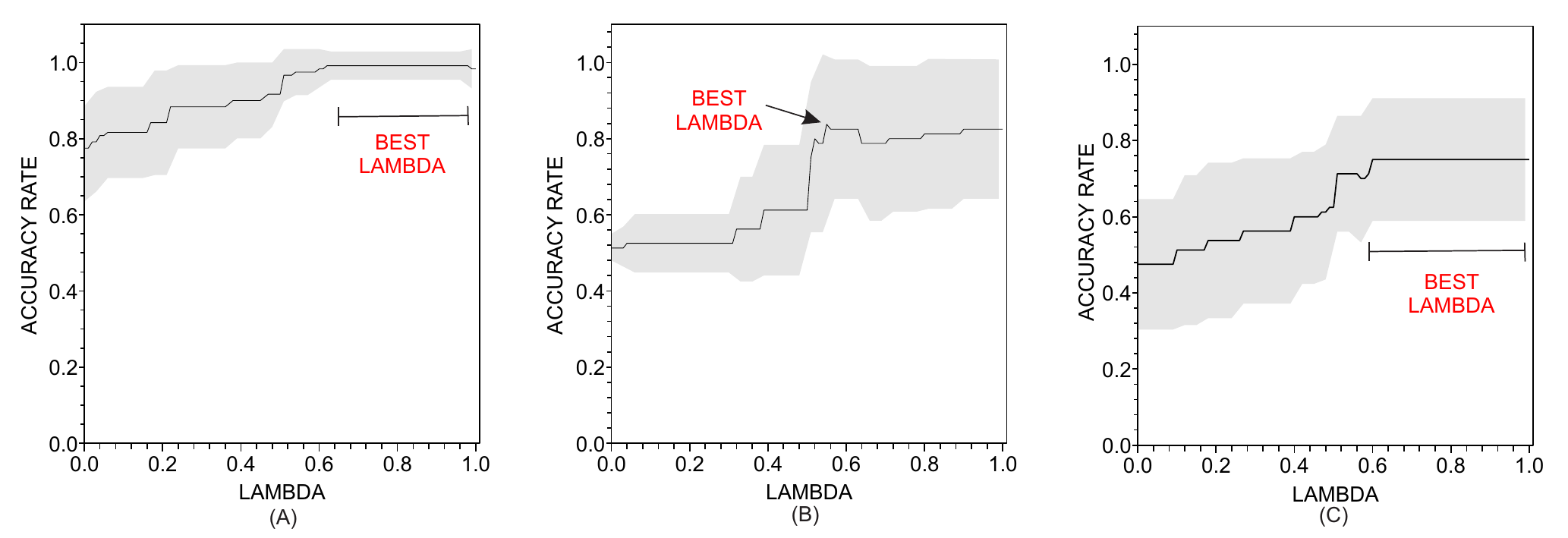}
    \end{center}
    \caption{\label{fig:acertim}Variation of the accuracy rate $\Gamma_H$ in function of $\lambda$ for a particular pair of homonymous authors in the (a) CN; (b) TI; and (c) LC databases. Note that in (a) and (b) the best $\Gamma_H$ includes $\lambda=1$. This means that the best accuracy \Red{can be obtained} exclusively with topological features in particular cases. The hybrid effect can be observed in (b) because there is a mixture of both traditional and topological approaches.}
\end{figure}

\begin{figure}
    \begin{center}
        \includegraphics[width=1\textwidth]{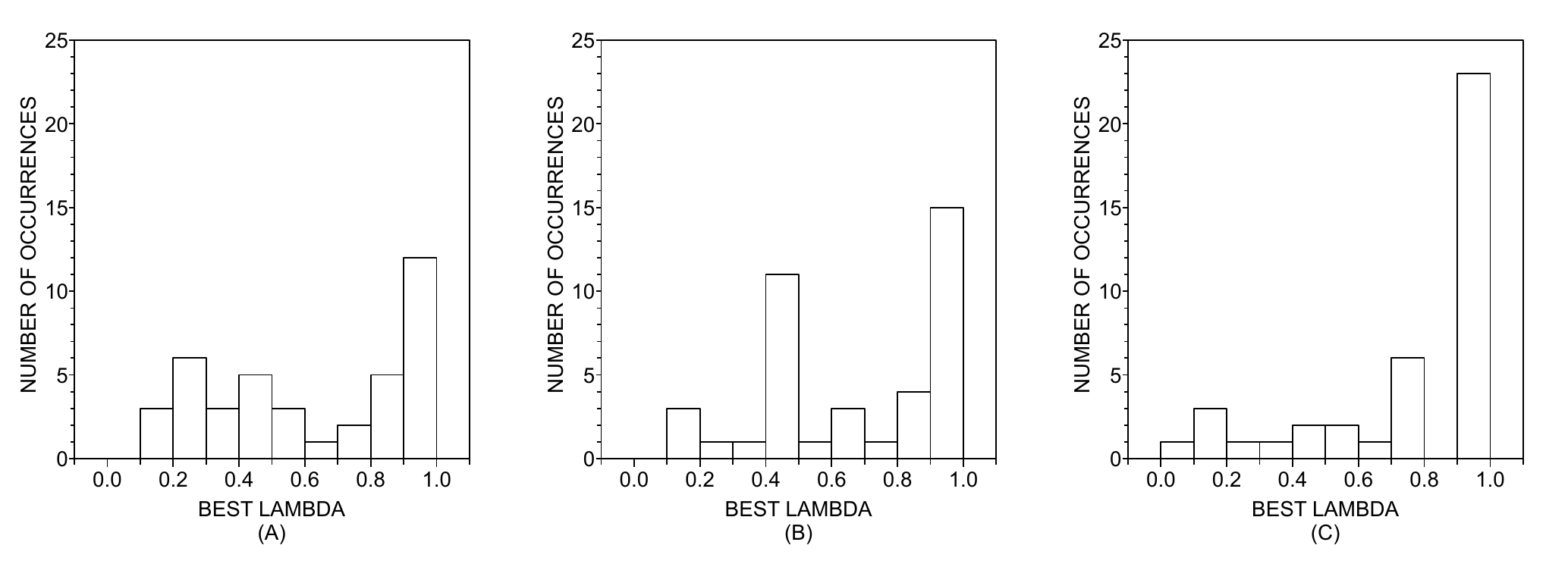}
    \end{center}
    \caption{\label{fig:dists} Distribution of $\lambda^*$ for the (a) CN ; (b) TI; and (c) LC databases. It is worth noting that $\lambda^* > 0$ \Red{in most cases}, which reinforce the suitability of topological measurements to improve the performance of the task.
    }
\end{figure}

The traditional approach based on the use of collaboration patterns was compared with the hybrid strategy as follows. For each pair of homonymous authors, we employed the hybrid algorithm described in Section \ref{sec:fuzzy} to take into account the topological information. The accuracy rate $\Gamma_H$ was computed for each $\lambda$ and then the best $\lambda$ (i.e. $\lambda^*$) was determined. Figure \ref{fig:acertim} illustrates the variation of $\Gamma_H$ as $\lambda$ varies within the interval $\lambda \in [0,1]$ for three specific pairs of homonymous authors. In all three cases, the best discrimination occurs with $\lambda^* > 0$, hence confirming that the use of topological features is useful to enhance the disambiguation task. In particular cases, it is even redundant to use the collaborative approach because the optimum $\Gamma_H$ was obtained with $\lambda = 1$ (see e.g. Figure \ref{fig:acertim}(a) and \ref{fig:acertim}(c)). In other cases, the best $\Gamma_H$ was observed through a convex combination of strategies, i.e. $0 < \lambda < 1$ (see Figure \ref{fig:acertim}(b)). The distribution of $\lambda^*$ for all pairs of authors displayed in Figure \ref{fig:dists} confirms that the topology plays a fundamental role on the task because in almost all databases $\lambda^* \neq 0$.



The individual accuracy rates, viz. $\Gamma_C$ (for the traditional collaborative approach) and $\Gamma_T$ (for the topological approach) were compared with the hybrid strategy in Table \ref{tab:literario}, whose lines represent \Red{a specific} pair of homonymous authors. For each database (CN, LC and TI), we show the $15$ best results where the hybrid approach displayed the highest increase in performance when compared with the traditional strategy.
In other words, the rows  are sorted in decreasing order of the margin of gain $\Gamma_{HC} = \Gamma_H - \Gamma_C$. As the table shows, the magnitude of $\Gamma_{HC}$  depends on the database. While for the CN and LC networks the maximum improvement in performance reached $32.5 \%$ and $30.0~\%$ respectively, for the TI database the maximum gain $\Gamma_{HC}$ did not exceed $10.0 \%$. These results suggest that the discrimination in the latter case is hampered because the TI collaborative network presents a more regular topological structure when compared with the other two networks.

\begin{table}[!ht]
\centering
\caption{Values of accuracy rate using the traditional ($\Gamma_C$), topological ($\Gamma_R$) and hybrid $\Gamma_H$ approaches. Rows are sorted by the margin of gain $\Gamma_{HC} = \Gamma_H - \Gamma_C$ in each database. The values of $\lambda$ concerning the best $\Gamma$ are also provided in the last columns. Note in particular that the margin of gain $\Gamma_{HC}$ vary from network to network, reaching a maximum of 32.5 \% in the CN database.}
\begin{tabular}{@{}cccc|cccc|cccc}
  \hline
  \multicolumn{4}{c|}{\bf Complex Networks} & \multicolumn{4}{c|}{ \bf Topological Insulators}  & \multicolumn{4}{c}{\bf Liquid crystals} \\
  \hline
  $\Gamma_C$  & $\Gamma_T$  & $\Gamma_H$ & $\lambda^*$  & $\Gamma_C$  & $\Gamma_T$  & $\Gamma_H$ & $\lambda^*$  & $\Gamma_C$  & $\Gamma_T$  & $\Gamma_H$ & $\lambda^*$  \\
  (\%) & (\%) & (\%) & -- & (\%) & (\%) & (\%) & -- & (\%) & (\%) & (\%) & -- \\
  \hline
  {51.3} & 82.5 & 83.8 & 0.55 & {88.8} & 98.8 & 98.8 & 1.00 & {52.5} & 82.5  & 82.5 & 1.00 \\
  {47.5} & 75.0 & 75.0 & 1.00 & {35.8} & 41.6 & 42.5 & 0.85 & {50.0} & 72.5  & 72.5 & 1.00 \\
  {77.5} & 98.3 & 99.2 & 0.98 & {62.5} & 63.8 & 68.8 & 0.56 & {67.5} & 77.5  & 85.0 & 0.77 \\
  {72.5} & 90.0 & 90.0 & 1.00 & {28.8} & 35.0 & 35.0 & 1.00 & {72.5} & 82.5  & 87.5 & 0.98 \\
  {81.3} & 97.5 & 97.5 & 1.00 & {18.1} & 20.6 & 23.8 & 0.88 & {57.5} & 71.3  & 71.3 & 1.00 \\
  {85.0} & 97.5 & 98.8 & 0.88 & {95.0} & 98.7 & 100.0 & 0.79 & {65.0} & 70.0  & 77.5 & 0.53 \\
  {86.3} & 96.3 & 100.0& 0.29& {18.8} & 23.8  & 23.8 & 1.00 & {83.8} & 85.0  & 95.0 & 0.60 \\
  {86.3} & 93.8 & 98.8 & 0.83 & {61.3} & 66.3 & 66.3 & 1.00 & {82.5} & 87.5  & 92.5 & 0.31 \\
  {83.8} & 95.0 & 96.3 & 0.42 & {90.8} & 90.8 & 95.0 & 0.70 & {77.5} & 72.5  & 87.5 & 0.48 \\
  {85.0} & 94.0 & 96.3 & 0.85 & {20.0} & 23.3 & 23.3 & 1.00 & {85.0} & 77.5  & 92.5 & 0.29 \\
  {85.0} & 95.0 & 95.0 & 1.00 & {97.5} & 92.5 & 100.0& 0.37 & {75.0} & 60.0  & 82.5 & 0.03 \\
  {85.0} & 87.5 & 95.0 & 0.49 & {31.3} & 32.5 & 33.8 & 0.84 & {95.0} & 100.0 & 100.0 & 1.00\\
  {85.0} & 84.2 & 93.4 & 0.49 & {28.8} & 31.3 & 31.3 & 1.00 & {40.0} & 45.0  & 45.0 & 1.00 \\
  {85.9} & 94.1 & 94.2 & 1.00 & {26.6} & 29.2 & 29.2 & 1.00 & {80.0} & 85.0  & 85.0 & 1.00 \\
  {72.5} & 76.3 & 80.0 & 0.49 & {33.8} & 36.3 & 36.3 & 1.00 & {91.3} & 95.0  & 95.0 & 1.00 \\
  \hline
\end{tabular}
\label{tab:literario}
\end{table}


\subsection{Relevance of topological measurements}

In the previous section, we have observed that topological measurements extracted from collaborative networks are useful to enhance the performance of disambiguating homonymous authors. In this section, we examine
which topological factors accounts for the improvement in the discrimination ability. To achieve this, we adopted a procedure to rank the relevance of the topological measurements. For each of the three databases, we selected the pairs of homonymous authors whose discriminability (revealed by the ``$\kappa$NN crisp'' algorithm) falls below a given threshold. We chose to select only the worst traditional classifiers to perform the assessment of relevance of topological measurements in order to focus the analysis on the situations where the structural approach is in fact responsible for the disambiguation. For each pair of homonymous authors we computed the relevance upon assessing the discriminability of every possible classifier constructed with a smaller combination of the original measurements. Thus, if $\mathcal{F}$ represents the total number of measurements, then $\sum_{i=0}^{\mathcal{F}} \mathcal{F}!/(\mathcal{F}-i)!i! = 2^\mathcal{F}$ classifiers are generated. With this procedure we aimed at identifying the most relevant measurements as those appearing among those combinations of classifiers with the \Red{highest accuracy rates}~\cite{interm}. The relevance was then quantified from the analysis of the matrix $\mathcal{M}$, with $2^{\mathcal{F}}$ rows and $\mathcal{F}$ columns, whose elements $m_{ij}$ are given by
\begin{equation}\label{eq.mavg}
    m_{ij} = \left\{
    \begin{array}{ll}
        1 & \textrm{ if the $i$-th best combination employed the $j$-th feature}, \\
        0 & \textrm{ otherwise. } \\
    \end{array}
    \right.
\end{equation}
Note that if a feature $j$ always appears among the best classifiers then
\begin{equation}
    f(x) = \sum_{i=1}^{x} m_{ij}, \ \ \{x \in \mathbb{N}^* | x \leq 2^\mathcal{F} \}
\end{equation}
will increase quickly. On the other hand, if $j$ tends to appear among the worst classifiers, $f(x)$ increases significantly only when $x$ takes high values. Thus, the relevance $r(j)$ of feature $j$ can be computed as the \Red{area underneath the curve} (AUC):
\begin{equation} \label{relevvance}
    r(f) = \int_{1}^{2^F} f(x) \mathrm{d}x = \sum_{i=1}^{2^F} \sum_{k=1}^{i-1} m_{kj} + \frac{1}{2} \sum_{i=1}^{2^F} m_{ij}.
\end{equation}
An example of quantification of relevance is provided in \ref{apenda}.
At last, the final rank of features in each database is established following the decreasing geometric mean of $r(j)$ over all pairs of homonymous authors. Note that this ordering scheme corresponds to the the likelihood of reaching, just by chance, a rank as good as the one obtained in all pairs of homonymous authors in each database. The results are depicted in Table \ref{tab.literario} and summarized below.

\begin{table}[!ht]
\centering
\caption{Rank (R) of relevance of measurements for the CN, TI and LC databases. Each pair of homonymous authors was evaluated using the quantity $r(j)$ (see eq. (\ref{relevvance})) and the final rank in each database was taken as the geometric average of $r(j)$ in each pair of homonymous author. In all three databases, the most prominent topological measurement was the average shortest path length.}
\begin{tabular}{cccc}
  \hline
  {\bf R} & {\bf CN} & {\bf TI} & {\bf LC} \\
  \hline
  \#1  & $l$  & $l$ & $l$ \\
  \#2  & $k^{(1)}_n$ & $C^{(1)}$ & $k^{(1)}$ \\
  \#3  & $\Delta s^{(1)}$ & $B$ & $\Delta k$ \\
  \#4  & $\Delta k^{(1)}$ & $\Delta k^{(1)}$ & $s^{(1)}$ \\
  \#5  & $s^{(1)}_n$     & $\Delta s^{(1)}$ & $s^{(1)}_n$ \\
  \#6  & $\mathcal{L}$   & $\Delta s^{(1)}$ & $B$ \\
  \#7  & $B$ & $k^{(1)}$ & $C^{(1)}$ \\
  \#8  & $k^{(1)}$ & $k^{(1)}_n$ & $\Delta s^{(1)}$ \\
  \#9  & $C^{(1)}$ & $s^{(1)}$ & $k^{(1)}_n$ \\
  \#10 & $s^{(1)}$ & $\mathcal{L}$ & $\mathcal{L}$ \\
  \hline
\end{tabular}
\label{tab.literario}
\end{table}

\begin{itemize}

  \item {\bf Average shortest path length ($l$)}: this measurement is the most prominent feature in all three databases. This probably occurs because the average shortest path length, unlikely other traditional collaborative features, takes into account long-range connections that might be essential to differentiate distinct patterns of organization occurring in deeper hierarchies.

  \item {\bf Degrees} ($k^{(1)}$, $k_n$ and $\Delta k_n$): these measurements are relevant only in particular cases. $k_n^{(1)}$ appears at the second position in the CN database, while the number of neighbors $k^{(1)}$ is the second most relevant feature in the LC database.
      Remarkably, for the TI database, $k^{(1)}$ and $k_n$ appear at the lower bottom of the relevance table probably because the  connections among neighbors play an essential role to discriminate ambiguous authors and this characteristic is only captured with the clustering coefficient (see discussion below). The standard deviation of the degree of neighbors $\Delta k_n$ does not seem to be relevant because it lies at intermediary positions in all three databases.

  \item {\bf Strength} ($s^{(1)}$, $s_n$ and $\Delta s_n$): the strength computed at the first levels are ranked into intermediary positions. The introduction of weights in edges does not seem to enhance significantly the discriminability because the differences $r(k^{(1)}) - r(s^{(1)})$ and $r(k_n) - r(s_n)$ are relatively small.

  \item {\bf Betweenness} ($B$): the number of shortest paths passing through nodes representing homonymous authors plays a minor role on the disambiguation task.  The best rank achieved was $r(B) = 3$rd position in the TI database. Actually, the weak performance of this quantity reveals that the length of paths and not their quantity is responsible for the improvement in the ability of discriminating authors' names.

  \item {\bf Clustering coefficient} ($C$): the connectivity among neighbors seems to be useful only for the TI database. This result contrasts with the result observed for the CN network, where the number of connections outgoing from neighbors was found to be relevant (note that $k_n$ was placed at the 2nd position in the CN rank). As one can observe in the TI rank, the total number of external links leaving neighbors is less informative for the discrimination task because $C^{(1)}$ is more important than other measurements that considers external connections such as $\Delta k$, $\Delta s$ and $k_n$. This means that, in the TI database, the feature responsible for the distinguishability of authors is the degree of connectivity among collaborators. In other words, homonymous authors might be differentiated from each other because they belong to communities with distinct modularity~\cite{modularidade}.

   \item {\bf Locality index} ($\mathcal{L}$): the percentage of connections located at the first hierarchical level turned out to be weakly discriminative for it achieved only the 6th position at the best case. This result suggests that the locality index of authors in collaborative networks does not vary across authors. Remarkably, even though the total {\bf number} of external ($k^{(1)}$) and internal links ($C^{(1)}$) is relevant in some cases, the {\bf proportion} of links located more closely to the node under analysis is not.

\end{itemize}

\subsection{Is it possible to predict $\lambda^*$?}

In the current investigation we found that the exact choice of the parameter $\lambda$ gives rise to a hybrid technique whose combination of strategies is able to enhance the performance of traditional methods relying solely on the recurrence of collaborations. The drawback associated with this methodology is that it cannot be straightforwardly employed in real applications because there is no a priori information available that could guide the selection of the best $\lambda$, i.e. $\lambda^*$.

\begin{figure}
    \begin{center}
        \includegraphics[width=1\textwidth]{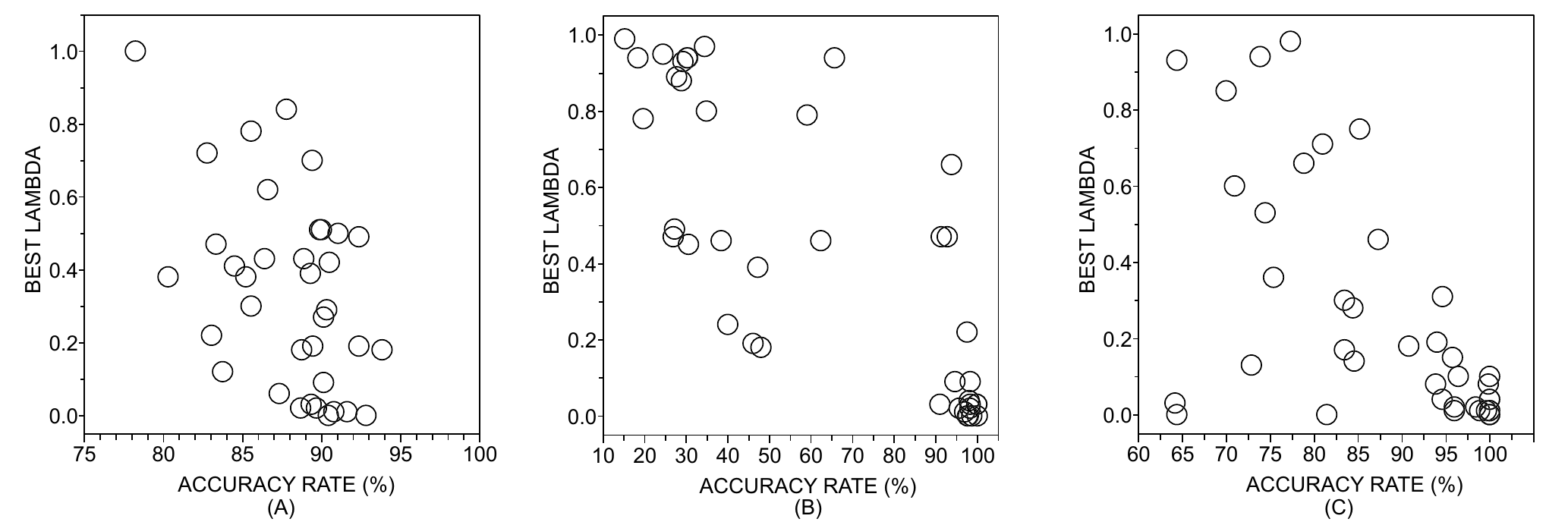}
    \end{center}
    \caption{\label{fig:correlations} Predicting the value of $\lambda^*$ from the analysis of discriminability achieved with the traditional collaborative approach. The best fitting obtained from the CN, TI and LC networks are: (a) $\lambda^* = 3.3222 - 3.389 \times 10^{-2}~\Gamma$ ($r = 0.46$ and $p = 4.8 \times 10^{-3}$; (b) $\lambda^* = 1.026 - 9.3 \times 10^{-3}~\Gamma$ ($r = 0.79$ and $p \simeq 0$) and (c) $\lambda^* = 1.68433 - 1.633 \times 10^{-2}~\Gamma$ ($r = 0.63$ and $p \simeq 0$).
    }
\end{figure}

The strategy we propose to estimate $\lambda^{*}$ is based on the assumption that the use of topological features become relevant especially when the discrimination achieved with collaborative attributes is unable to provide a reasonable discrimination. More specifically, our strategy assigns the highest values of $\lambda^{*}$ to the worst traditional classifiers in the following manner. Initially, we evaluate the discriminability of homonymous authors by computing the accuracy rate $\Gamma_c$ obtained with collaborative features. Next, the value of $\lambda^{*}$ is identified using the hybrid approach according to eq. (\ref{lambida}). Finally, $\lambda^{*}$ is correlated with $\Gamma_C$ and the best line fitting the data is extracted.
In some cases it was imperative to choose a particular value of $\lambda^*$ because the best $\lambda$ may occur in intervals (see e.g. Figure \ref{fig:correlations}(a) and (c)). In order to construct the regression model maximizing its prediction ability we chose, in these cases, the value of $\lambda$ maximizing the correlation between $\Gamma_C$ and $\lambda^*$ with the simulated annealing heuristic~\cite{duda}. The relationship between $\Gamma_C$ and $\lambda^*$ and the best curve fitting the data are displayed in Figure \ref{fig:correlations}. Note that, in all three databases, the correlation between $\Gamma_C$ and $\lambda^*$ reached significant values ($r = 0.46$, $r = 0.79$ and $r = 0.63$ respectively for CN, TI and LC), suggesting that the optimal weighting for the hybrid approach can be reasonably predicted from the traditional analysis of collaborations.

\section{Conclusions} \label{conclusao}

The problem of scientific collaboration has been studied by a large number of works.
However, the usefulness of topological measurements of complex networks for discriminating authors' names has not been usually taken into account. Actually, the study of topological features of collaborative networks for discriminating homonymous authors has been restricted to a few works~\cite{amancio1}. In the present paper, we devised a classification scheme which considers synchronously both traditional and topological strategies for characterizing authors in scientific papers, {which allowed us to obtain an improved version of the collaborative disambiguation strategy.} This classification scheme was accomplished by defining a classifier whose final inference is based on a weighted convex combination of decisions based on semantical and topological patterns of collaboration. Several interesting results have been obtained. First, the weight related to the topological strategy was different from zero in most cases, implying that the topological approach plays a complementary role on the characterization of homonymous authors. Another interesting finding was that when researchers are poorly distinguished by the analysis of recurrence of co-authorship patterns, the average distance between authors in a given research area was the main topological feature providing valuable information about authors' peculiarities. We also verified the correlation between the weight associated with the topological strategy and the best accuracy rate of the disambiguating system. It followed from this result that the weights are inversely proportional to the distinguishability achieved with co-authorship patterns. As such, the best weight related to the topological approach could be predicted automatically. Further works could introduce a decay factor for edges weights so that old collaborations are mapped into weaken edges. Another possibility is to take into account as well the organization of the authors' collaboration in the attribute space~\cite{highorder}.

\begin{acknowledgements}
The authors acknowledge the financial support from {S\~ao Paulo Research Foundation (FAPESP)} (grant numbers 2010/00927-9, 2011/50761-2 and 2013/06717-4).
\end{acknowledgements}



\newpage

\appendix
\section{Quantifying feature relevance} \label{apenda}

The method employed to estimate the relevance of features is based on the frequency of its appearance among the most accurate classifiers, when one considers all the possible $2^{\mathcal{F}}$ combinations of attributes. The fraction of classifiers displaying an accuracy rate above a specific threshold is calculated and the area underneath the curve obtained by varying the threshold in decreasing order is employed as a relevance estimator. This methodology has been chosen because it draws on multivariate analysis, which provides a powerful ability for representing and analyzing non-trivial relationships among many features. More simple methods, such as the Kullback-Leibler divergence~\cite{duda}, are unable to grasp complex interactions between features.

\begin{figure*}[!ht]
    \begin{center}
        \includegraphics[width=1\textwidth]{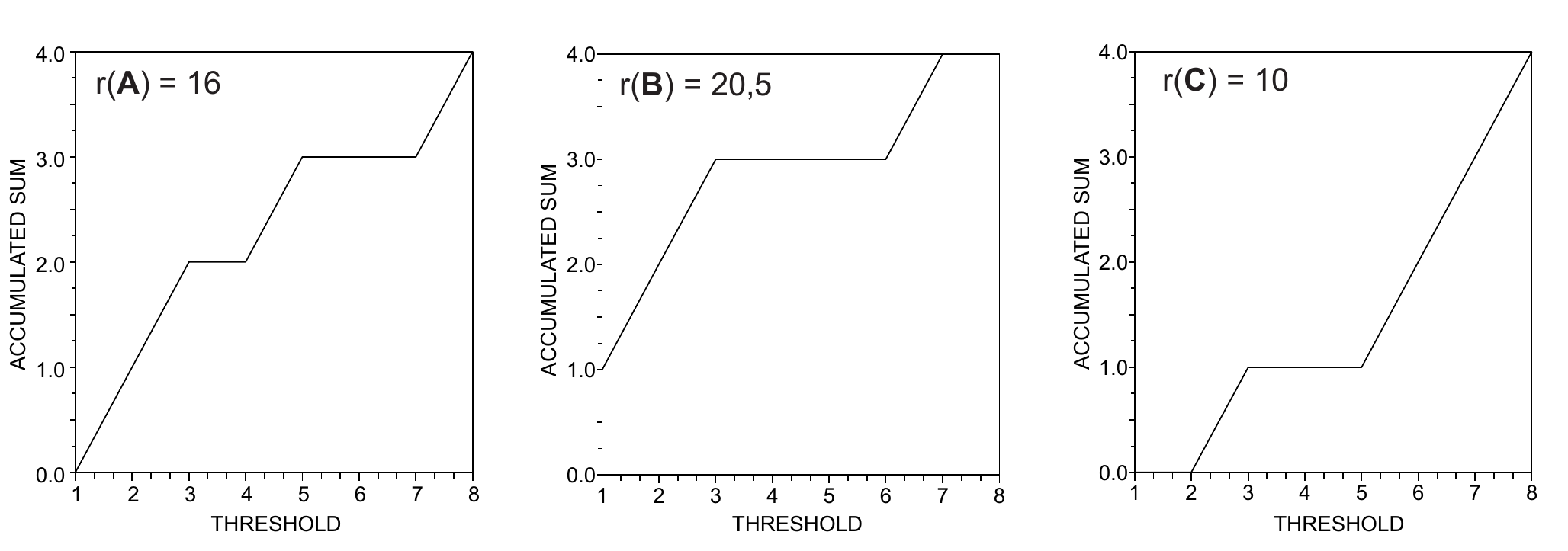}
    \end{center}
    \caption{\label{fig:ambs} Quantification of relevance of the attributes
    (a) $\mathbf{A}$, (b) $\mathbf{B}$ and (c) $\mathbf{C}$ using eq. (\ref{relevvance}). The assumed order of performance of classifiers is given in Table \ref{tab.exemplo}. Note that attribute $\mathbf{B}$ grows more rapidly because it is employed in the best classifiers. On the other hand, the attribute $\mathbf{C}$ is the less relevant for it appears predominantly among the classifiers with the worst performance in Table \ref{tab.exemplo}.}
\end{figure*}

Table \ref{tab.exemplo} exemplifies the quantification of relevance of three features (viz. $\mathbf{A}$, $\mathbf{B}$ and $\mathbf{C}$) describing a toy database according to eq.(\ref{relevvance}). The combinations of classifiers are sorted in decreasing order of precision. For example, the classifier induced solely with the attribute $\mathbf{B}$ displayed the best performance while the classifier induced with $\mathbf{A}$ and $\mathbf{C}$ achieved the second highest accuracy rate. The curves displaying the accumulated frequency above the established threshold for each attribute is given Figure \ref{fig:ambs}.


\begin{table*}[h]
    \centering
    \caption{\label{tab.exemplo}
    Example of computation of relevance $r$ of features according to eq. (\ref{relevvance}). We consider a toy database represented by three attributes, viz. $\mathbf{A}$, $\mathbf{B}$ and $\mathbf{C}$. The classifiers are sorted in decreasing order of precision and the attributes employed in each of the $2^3 = 8$ combinations of classifiers are marked in the respective column with a dot ($\bullet$). Note that the most efficient classifiers usually employ attribute $\mathbf{B}$. Since $\mathbf{B}$ is used frequently to induce the best classifiers, its respective curve presents the fastest growing (see Figure \ref{fig:ambs}). As a result, $r(\mathbf{B}) > r(\mathbf{A})$ and $r(\mathbf{B}) > r(\mathbf{C})$.
    }
    \begin{tabular}{|c|c|c|c|c|c|c|c|c|}
        \hline
        {\bf Rank} & \#1 & \#2 & \#3 & \#4 & \#5 & \#6 & \#7 & \#8 \\
        \hline
        Attribute $\mathbf{A}$ & & $\bullet$ & $\bullet$ & & $\bullet$ & & & $\bullet$ \\
        Attribute $\mathbf{B}$ & $\bullet$ & $\bullet$ & $\bullet$ & & & & $\bullet$ & \\
        Attribute $\mathbf{C}$ & & & $\bullet$ &  & & $\bullet$ & $\bullet$ & $\bullet$ \\
        \hline
        \end{tabular}
    \end{table*}

\end{document}